\documentclass[conference]{IEEEtran}
\renewcommand{\baselinestretch}{0.94}

\usepackage{cite}
\usepackage{acronym}
\usepackage{amsmath,amssymb,amsfonts}
\usepackage{cleveref}
\usepackage{bm}
\usepackage{nicefrac}
\usepackage{siunitx}
\usepackage{graphicx}
\usepackage{textcomp}
\usepackage{xcolor}
\usepackage{caption}
\usepackage{subcaption}
\usepackage{cite}
\usepackage{url}
\usepackage{nicefrac}
    
\usepackage{tikz}
\usetikzlibrary{math}
\usetikzlibrary{shapes.geometric,calc}
\usetikzlibrary{arrows.meta, quotes, angles}
\usetikzlibrary{decorations}
\usetikzlibrary{chains,shapes.misc,positioning,scopes}
\usepackage{circuitikz}
\usetikzlibrary{circuits.ee.IEC}
\usetikzlibrary{decorations}

\usetikzlibrary{shadows}
\usepackage{subcaption}

\acrodef{AMPAR}{$\alpha$-amino-3-hydroxy-5-methyl-4-isoxazolepropionic acid receptor}
\acrodef{MC}{molecular communication}
\acrodef{nAChR}{nicotinic acetylcholine receptor}
\acrodef{NMDAR}{$N$-methyl-\textsc{D}-aspartate receptor}
\acrodef{NT}{neurotransmitter}
\acrodef{PBS}{particle-based simulation}
\acrodef{Rx}{receiver}
\acrodef{SMC}{synaptic molecular communication}
\acrodef{SSD}{state-space description}
\acrodef{TFM}{transfer function model}
\acrodef{Tx}{transmitter}
\acrodef{wrt}[w.r.t.]{with respect to}

\newcommand{\kap}[1]{\kappa_{\mathrm{#1}}}
\newcommand{\kapt}[1]{\tilde{\kappa}_{\mathrm{#1}}}
\newcommand{\ke}{\kap{e}}
\newcommand{\kcd}{\kap{cd}}
\newcommand{\kco}{\kap{co}}
\newcommand{\kdo}{\kap{do}}
\newcommand{\kdc}{\kap{dc}}
\newcommand{\koc}{\kap{oc}}
\newcommand{\kod}{\kap{od}}

\newcommand{\kcdt}{\kapt{cd}}
\newcommand{\kcot}{\kapt{co}}
\newcommand{\kdot}{\kapt{do}}
\newcommand{\kdct}{\kapt{dc}}
\newcommand{\koct}{\kapt{oc}}
\newcommand{\kodt}{\kapt{od}}

\newcommand{\cstate}{\bar{\mathrm{C}}}
\newcommand{\ostate}{\bar{\mathrm{O}}}
\newcommand{\dstate}{\bar{\mathrm{D}}}

\newcommand{\As}{\bm{\mathcal{A}}}

\newcommand{\tran}{^{\scriptscriptstyle\mathrm{T}}}
\newcommand{\tra}{{\scriptscriptstyle\mathrm{T}}}

\makeatletter
\long\def\@makecaption#1#2{\ifx\@captype\@IEEEtablestring%
    \footnotesize\begin{center}{\normalfont\footnotesize #1}\\
        {\normalfont\footnotesize\scshape #2}\end{center}%
    \@IEEEtablecaptionsepspace
    \else
    \@IEEEfigurecaptionsepspace
    \setbox\@tempboxa\hbox{\normalfont\footnotesize {#1.}~~ #2}%
    \ifdim \wd\@tempboxa >\hsize%
    \setbox\@tempboxa\hbox{\normalfont\footnotesize {#1.}~~ }%
    \parbox[t]{\hsize}{\normalfont\footnotesize \noindent\unhbox\@tempboxa#2}%
    \else
    \hbox to\hsize{\normalfont\footnotesize\hfil\box\@tempboxa\hfil}\fi\fi}
\makeatother

\begin{document}
%
\title{Signal Reception With Generic Three-State Receptors in Synaptic MC
\vspace*{-0.5cm}
}

\author{\IEEEauthorblockN{Sebastian Lotter\textsuperscript{*}, Michael T. Barros\textsuperscript{\textdagger}, Robert Schober\textsuperscript{*}, and Maximilian Schäfer\textsuperscript{*}}
\IEEEauthorblockA{\textit{\textsuperscript{*}Friedrich-Alexander-Universität Erlangen-N\"urnberg (FAU),} Erlangen, Germany 
\\
	\textit{\textsuperscript{\textdagger}School of Computer Science and Electronic Engineering, University of Essex,}
	Colchester, UK}
	\vspace*{-1.0cm}
}

\maketitle

\begin{abstract}
Synaptic communication is studied by communication engineers for two main reasons.
One is to enable novel neuroengineering applications that require interfacing with neurons.
The other reason is to draw inspiration for the design of synthetic \acl{MC} systems.
Both of these goals require understanding of how the chemical synaptic signal is sensed and transduced at the synaptic \ac{Rx}.
While signal reception in \ac{SMC} depends heavily on the kinetics of the receptors employed by the synaptic \acp{Rx}, existing channel models for \ac{SMC} either oversimplify the receptor kinetics or employ complex, high-dimensional kinetic schemes limited to specific types of receptors.
Both approaches do not facilitate a comparative analysis of different types of natural synapses.
In this paper, we propose a novel deterministic channel model for \ac{SMC} which employs a generic three-state receptor model that captures the characteristics of the most important receptor types in \ac{SMC}.
The model is based on a transfer function expansion of Fick's diffusion equation and accounts for release, diffusion, and degradation of \aclp{NT} as well as their reversible binding to finitely many generic postsynaptic receptors.
The proposed \ac{SMC} model is the first that allows studying the impact of the characteristic dynamics of the main postsynaptic receptor types on synaptic signal transmission.
Numerical results indicate that the proposed model indeed exhibits a wide range of biologically plausible dynamics when specialized to specific natural receptor types.
\end{abstract}

\acresetall
\section{Introduction}
\label{sec:intro}

\Ac{MC} is an emerging research area employing molecules as information carriers in the quest for future in-body communication solutions \cite{Akyildiz2015TheThings}.
As a subfield of \ac{MC}, \ac{SMC} is concerned with \ac{MC} in chemical synapses, where information is conveyed between cells by chemical messengers called {\em \acp{NT}}.
One of the driving forces behind \ac{SMC} research is the development of novel neuroengineering applications \cite{Veletic2019SynapticInterfaces,Akan2021InformationChallenges}.
In addition, \ac{SMC} is a natural \ac{MC} system that has evolved over thousands of years to cope with the challenges and requirements imposed by its natural environment, the human body.
Hence, it can also present a blueprint for the development of synthetic \ac{MC} systems.
However, both these goals require a thorough understanding of how efficient and reliable communication is facilitated in the synapse by its specific molecular machinery.
%
%

\begin{figure}[t]
    \centering
    \includegraphics[width=\linewidth]{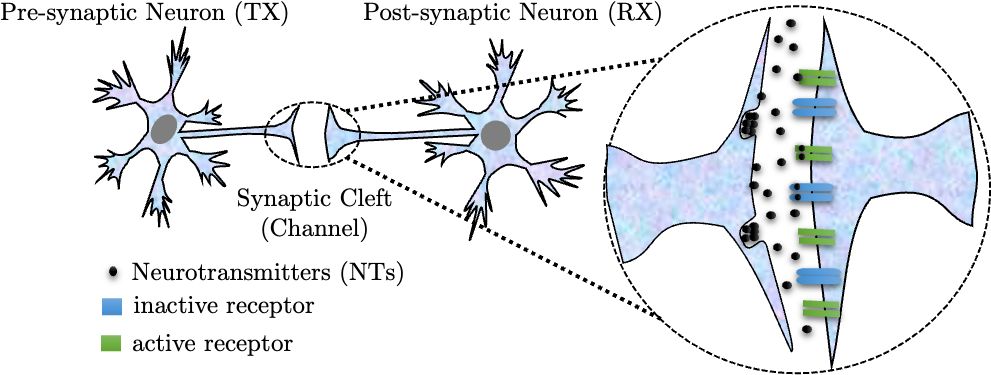}
    \vspace*{-0.7cm}
    \caption{Neurotransmitters (black) are released into the synaptic cleft from the pre-synaptic neuron and bind to receptors (green: active receptor, blue: inactive receptor) at the post-synaptic neuron.}
    \label{fig:scenario}
    \vspace*{-0.6cm}
\end{figure}

In \ac{SMC}, \acp{NT} released by the exocytosis of intracellular vesicles from a {\em presynaptic cell}, the \ac{Tx}, propagate inside the {\em synaptic cleft} (channel) towards a {\em postsynaptic cell}, the \ac{Rx}, cf.~Fig.~\ref{fig:scenario}.
At the postsynaptic cell, \acp{NT} bind reversibly to transmembrane receptors which transduce the synaptic signal into a postsynaptic intracellular downstream signal \cite[Chap.~10]{MolToNet}.
There exists a huge variety of postsynaptic receptor types in nature and the way in which the postsynaptic cell parses a given synaptic signal depends largely on which receptor types are used for signal transduction.
Accordingly, different types of postsynaptic receptors mediate complementary functions inside the postsynaptic cell
\cite[Chap.~16]{MolToNet}.
Hence, understanding the communication theoretic design of \ac{SMC} requires comprehensive channel models that capture the functional heterogeneity of postsynaptic signal transduction.
At the same time, these models should provide enough generality to draw conclusions on the fundamental design implications of different receptor types.

Existing works have studied numerous aspects of \ac{SMC} using tools from information and communication theory.
Physical channel models for \ac{SMC} based on the diffusion equation were developed in \cite{Khan2017Diffusion-BasedChannel,Bilgin2017ASynapse,Lotter2021SynapticSynapse,Lotter2021SaturatingModels}.
Synaptic communication was also studied from a communication theoretic perspective in \cite{Balevi2013ACommunications} and in terms of its fundamental information theoretic limits in \cite{Veletic2016OnSynapses,Veletic2020AnChannels}.
The \ac{SMC} models that explicitly consider the binding of individual \acp{NT} to postsynaptic receptors either assume a two-state (closed/open) kinetic scheme\footnote{We note that the term \textit{kinetic scheme} is used equivalently to the term \textit{Markov kinetic model} as explained in \cite{Destexhe1994SynthesisFormalism}.} for the receptors \cite{Khan2017Diffusion-BasedChannel,Lotter2021SynapticSynapse,Lotter2021SaturatingModels,Veletic2016OnSynapses,Veletic2020AnChannels} or adopt intricate multi-state models with many degrees of freedom \cite{Bilgin2017ASynapse}.
However, the commonly employed two-state models neglect the desensitization of receptors, which is an important property of the main postsynaptic receptor types, see Fig.~\ref{fig:receptorFull}, and experimental data suggests that kinetic schemes with at least three states are required to reproduce the variety of receptor responses to \ac{NT} releases observed in nature \cite{Destexhe1994SynthesisFormalism}.
In order to deduce fundamental insights on the impact of different receptor types on system design, communication theoretic analysis has to rely on tractable models with limited degrees of freedom and sufficient generality.
However, the level of detail of existing \ac{SMC} signal models is not sufficient for understanding the implications that different receptor kinetics have on the signal reception.

\Ac{SMC} \acp{Rx} belong to the class of {\em reactive \acp{Rx}} which have been studied as one of the generic \ac{Rx} models in \ac{MC} \cite{jamali:ieee:2019}.
However, with the notable exception of the models proposed in \cite{Chou2015ImpactNetworks,Awan2017MolecularNetworks}, chemical reaction networks at the \ac{Rx} with more than two states have not been investigated thoroughly yet \cite{jamali:ieee:2019}.
The mesoscopic, i.e., voxel-based, models proposed in \cite{Chou2015ImpactNetworks,Awan2017MolecularNetworks} show that, in a rather general \ac{MC} setting, the performance can benefit from multi-state \ac{Rx} kinetics.
While these models are not specific to \ac{SMC}, they provide additional motivation to study the properties of \ac{SMC} \acp{Rx} to better understand their implications for communication performance.

In this paper, we propose a novel deterministic signal model for \ac{SMC} incorporating the diffusion and degradation of \acp{NT} inside the synaptic cleft and their reversible binding to different types of postsynaptic receptors.
The proposed model is based on the expansion of the diffusion equation in terms of transfer functions and incorporates the receptor kinetics as feedback loops in a \ac{SSD}.
It extends the model developed by the authors in \cite{Lotter2021SaturatingModels} by a generic three-state receptor which encompasses the characteristics of the main postsynaptic receptor types found in nature.
The proposed model allows for the first time to analyze the impact of different receptor types on postsynaptic signal transduction.
Dimensional analysis is conducted to ensure the generality of the proposed model and three-dimensional stochastic \acp{PBS} confirm its accuracy.
Numerical results obtained by specializing the proposed generic model to \acp{AMPAR} and \acp{NMDAR}, respectively, suggest that a rich set of synaptic dynamics observed in natural synapses and specific to the respective receptor type, can be reproduced.
At the same time, the analysis remains tractable due to the relatively small number of model parameters.
Hence, we are confident that the generic \ac{Rx} model proposed in this paper can contribute to the understanding of signal transduction in \ac{SMC} and inspire the design of synthetic \ac{MC} systems based on reactive \acp{Rx}.

The paper is organized as follows.
In Section~\ref{sec:system}, we introduce the considered \ac{SMC} system and conduct the dimensional analysis.
In Section~\ref{sec:tfm}, the deterministic signal model 
is developed.
Finally, Section~\ref{sec:simulation} presents numerical results and our main conclusions are summarized in Section~\ref{sec:conclusion}.

\section{System Model}
\label{sec:system}
\vspace*{-0.1cm}
\subsection{Biological Background}
\vspace*{-0.1cm}
In \ac{SMC}, the presynaptic cell secretes \acp{NT} which propagate by Brownian motion through the synaptic cleft and bind reversibly to cell-surface receptors at the postsynaptic cell.
The activation of the postsynaptic receptors triggers an electrochemical downstream signal which propagates inside the postsynaptic cell.
To terminate synaptic transmission, \acp{NT} are removed from the synapse either by \ac{NT} uptake at glial cells or at the presynaptic cell \cite{Danbolt2001GlutamateUptake}, respectively, or by enzymatic degradation 
\cite[Chap.~15]{MolToNet}.
In this paper, the focus is on exploring the transduction of the chemical synaptic signal by the postsynaptic receptors.

Depending on the type of synapse, different types of postsynaptic receptors can be found in \ac{SMC} 
\cite[Chap.~10]{MolToNet}.
They mediate different types of postsynaptic downstream signals in reponse to the presynaptic release of \acp{NT}.
Postsynaptic receptors are broadly classified into {\em ionotropic} receptors, i.e., ligand-gated ion channels, and {\em metabotropic} receptors, i.e., receptors that activate intracellular second messengers.
However, even among the different receptor subtypes within the class of ionotropic receptors exist considerable differences in terms of their functionality.
The most prominent ionotropic receptors found in human synapses are \acp{AMPAR}, \acp{NMDAR}, and \acp{nAChR}, where the first two are expressed in glutamatergic synapses and the latter one is expressed in cholinergic synapses.
Of the two ionotropic glutamate receptors, \acp{AMPAR} are known to contribute a fast-rising component to the postsynaptic downstream signal and mediate the postsynaptic response to single synaptic transmission events.
\Acp{NMDAR} in turn contribute a late component to the postsynaptic downstream signal 
\cite[Chap.~16]{MolToNet}.

\begin{figure}[t]
    \centering
    \includegraphics[width=0.75\linewidth]{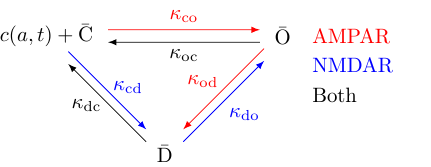}
    \vspace*{-0.3cm}
    \caption{Generic three-state receptor model proposed in \cite{Destexhe1994SynthesisFormalism}. $\cstate$, $\ostate$, $\dstate$: Closed, open, and desensitized receptor state, respectively. $c(a,t)$: Concentration of \acp{NT} applied to postynaptic receptors. State transitions specific for \acp{AMPAR}, \acp{NMDAR}, and both receptors are shown in red, blue, and black, respectively.}
    \label{fig:receptorFull}
    \vspace*{-0.6cm}
\end{figure}
\vspace*{-0.1cm}
\subsection{Generic Three-State Receptor Model}
\label{subsec:genRec}
\vspace*{-0.1cm}
The detailed analysis of the different receptor types in \ac{SMC} is complicated by the fact that the kinetic schemes for these receptor types available in the biology and neuroscience literature are rather intricate.
For example, the \ac{AMPAR} model proposed in \cite{Haausser1997DendriticCells} comprises $9$ different states and $22$ independent and non-zero rate constants.
The \ac{AMPAR} model proposed in \cite{Robert2003HowOccupancy} entails $16$ different states and $11$ independent non-zero parameters.
Clearly, the high-dimensional parameter spaces associated with these receptor models lead to problems in terms of model identifiability, i.e., the parameters of the model can not be reliably determined from experimental observations.
Moreover, the analysis of such high-dimensional models is cumbersome and often of limited generality, since even different subtypes of the same receptor type may have differing kinetic schemes.
To overcome these difficulties, the gating of different postsynaptic receptor types was analyzed on a macroscopic level in \cite{Destexhe1994SynthesisFormalism}.
Based on experimental measurements of postsynaptic currents, the authors in \cite{Destexhe1994SynthesisFormalism} identify kinetic schemes with reduced parameter spaces as compared to the models proposed in \cite{Haausser1997DendriticCells,Robert2003HowOccupancy} for the main types of postsynaptic receptors.
More specifically, the kinetic schemes proposed in \cite{Destexhe1994SynthesisFormalism} are derived from a generic kinetic scheme which is constrained to three different states and $6$ rate constants.
The three considered states correspond to the {\em closed}, {\em open}, and {\em desensitized} states.
In this model, a receptor in the closed state is {\em unbound} and {\em impermeable} to ions.
When the receptor is open, it is {\em bound} by one \ac{NT} and {\em permeable} to ions.
When the receptor is desensitized, it is {\em bound} by one \ac{NT} and {\em impermeable} to ions.
The $6$ rate constants correspond to the state transitions between the three possible states.
Depending on the specific type of receptor, some of the rates are zero.
Fig.~\ref{fig:receptorFull} shows the generic three-state receptor model from which the kinetic schemes of the different types of postsynaptic receptors were derived in \cite{Destexhe1994SynthesisFormalism}.
Despite their simplicity, the kinetic schemes proposed in \cite{Destexhe1994SynthesisFormalism} were shown to capture the main characteristics of the considered receptor types in terms of the postsynaptic current they mediate.
In the following sections, we couple the generic three-state receptor model proposed in \cite{Destexhe1994SynthesisFormalism} with the reaction-diffusion process governing the propagation and degradation of \acp{NT} in the synaptic cleft. 
The proposed model can be specialized to any type of postsynaptic receptor considered in \cite{Destexhe1994SynthesisFormalism} by adapting the reaction rate constants of the generic model accordingly.
When specialized to \acp{AMPAR} and \acp{NMDAR}, respectively, the proposed model shows richer dynamics than the simple two-state (closed/open) kinetic scheme.

\vspace*{-0.1cm}
\subsection{Dimensional Analysis}


The generic receptor model allows for rich receptor dynamics, resembling the characteristics of the main postsynaptic receptors in \ac{SMC}, while keeping the dimension of the parameter space of the model low.
In other words, the model is widely applicable and requires few parameters.
To achieve a similar level of generality for the yet to be derived model of the reaction-diffusion process, we first conduct dimensional analysis and then present the final model in dimensionless form.
This approach ensures on the one hand that the proposed model can be easily adapted to other parameter ranges than the ones considered in this paper.
On the other hand, by re-scaling the independent variables in the model, the total number of model parameters is reduced.

Let $c(x,t)$ be the concentration of \acp{NT} in the synaptic cleft in \si{\per\meter} at time $t$ and space $x$.
Under the assumptions stated and discussed in detail in \cite[Sec.~II]{Lotter2021SaturatingModels}, by Fick's second law, and in the presence of enzymatic degradation, $c(x,t)$ is governed by the following reaction-diffusion equation \cite[Eq.~(3)]{Lotter2021SaturatingModels}
\vspace*{-0.1cm}
\begin{equation}
    \delta_t c(x,t) = \!D \delta_{xx} c(x,t) - \ke c(x,t) + s(x,t), \!\!\quad 0 < x < a,
\end{equation}
where $\delta_t$ and $\delta_{xx}$ denote the first derivative \ac{wrt} $t$ and the second derivative \ac{wrt} $x$, respectively, $D$ denotes the diffusion coefficient of the \acp{NT} in \si{\square\meter\per\second}, $\ke$ denotes the enzymatic degradation rate in \si{\per\second}, $s(x,t)$ denotes the source term modeling the presynaptic relase of \acp{NT} in \si{\per\meter\per\second}, and $a$ denotes the width of the synaptic cleft in \si{\meter}.

Now, we switch to dimensionless variables before we proceed.
To this end, we define dimensionless time, dimensionless space, and the dimensionless concentration of \acp{NT} as
\vspace*{-0.1cm}
\begin{align}
    t'= \nicefrac{t\cdot D}{a^2}, \quad x'=\nicefrac{x}{a}, \textrm{ and } c'(x',t') = \nicefrac{c(x,t)\cdot a}{N_0},
\end{align}
respectively, where $N_0$ denotes the number of released molecules per vesicle.
With these substitutions, we obtain
\vspace*{-0.1cm}
\begin{equation}
    \delta_{t'} c'(x',t') = \delta_{x'x'} c'(x',t') - \ke' c'(x',t') + s'(x',t'),
    \label{eq:dl:pde}
\end{equation}
where $\ke' = \frac{\ke a^2}{D}$ denotes the dimensionless enzymatic degradation rate and $s'(x',t')$ is defined as follows
\vspace*{-0.1cm}
\begin{equation}
    s'(x',t')=\nicefrac{s(x,t)\cdot a}{N_0}.
\end{equation}
Furthermore, defining the dimensionless flux of \acp{NT} in $x'$-direction, $i'_x$, as follows
\vspace*{-0.2cm}
\begin{equation}
    i'_{x'}(\xi,t') = -\delta_{x'} c'(x',t')\big|_{x'=\xi},
\end{equation}
we obtain for the reflective left boundary of the synapse
\begin{equation}
    i'_{x'}(0,t') = 0,\label{eq:bc:left}
\end{equation}
and for the right boundary, which is covered by receptors,
\begin{equation}
    i'_{x'}(1,t') = \delta_{t'} o'(t') + \delta_{t'} d'(t'),
    \label{eq:bc:full}
\end{equation}
where $o'(t')$ and $d'(t')$ denote the dimensionless numbers of open and desensitized receptors, respectively.
$i'_{x'}(x',t')$ is related to its dimensional counterpart $i_x(x,t)$ as follows
\begin{equation}
    i_x(\zeta,t) = -D \delta_x c(x,t) \big|_{x=\zeta} = - \nicefrac{N_0 D}{a^2}\cdot i'_{x'}(\zeta/a,t').
\end{equation}
Adopting the three-state kinetic scheme introduced in Section~\ref{subsec:genRec}, we obtain
\begin{align}
    \delta_{t'} o'(t') &= \kco'\left(1 - \nicefrac{N_0}{C^*}\left[o'(t') + d'(t') \right]\right)c'(1,t')\nonumber\\
    &{}- \koc' o'(t') - \kod' o'(t') + \kdo' d'(t'),
    \label{eq:bc:open}
\end{align}
where $\kco' = \kco a/D$, $\koc' = \koc a^2/D$, $\kod' = \kod a^2/D$, $\kdo' = \kdo a^2/D$, and $C^*$ denote the dimensionless state transition rates (see Fig.~\ref{fig:receptorFull}), and the total number of postsynaptic receptors, respectively.
Similarly, we have
\begin{align}
    \delta_{t'} d'(t') &= \kcd'\left(1 - \nicefrac{N_0}{C^*}\left[o'(t') + d'(t') \right]\right)c'(1,t')\nonumber\\
    &{}- \kdc' d'(t') - \kdo' d'(t') + \kod' o'(t'),
    \label{eq:bc:desen}
\end{align}
with $\kcd' = \kcd a/D$, $\kdc' = \kdc a^2/D$, $\kdo' = \kdo a^2/D$, and $\kod' = \kod a^2/D$.
$o'(t')$ and $d'(t')$ are related to their respective dimensional counterparts $o(t)$ and $d(t)$ as follows
\begin{equation}
    N_0 o'(t') = o(t) \quad \textrm{ and } \quad N_0 d'(t') = d(t).
\end{equation}
Finally, assuming instantaneous release of \acp{NT} at $x=0$, $s'(x',t')$ derives from \cite[Eq.~(2)]{Lotter2021SaturatingModels} as follows
\vspace*{-0.1cm}
\begin{align}
    s'(x',t') \!=\!\! \sum_{m \in \mathcal{M}}\!\! \delta(t'-T_m')\delta(x'),
\end{align}
where $\delta(\cdot)$ denotes the Dirac delta distribution, $T_m$ denotes the \ac{NT} release times in \si{\second}, $\mathcal{M}$ denotes the set of release time indices, and $T_m'=T_m\frac{D}{a^2}$. 
All dependent variables, i.e., $o'(t')$, $d'(t')$, and $c'(x',t')$, are assumed to be zero at $t'=0$.

\section{Transfer Function Model}
\label{sec:tfm}
\vspace*{-0.1cm}
In this section, we derive a deterministic signal model for \ac{SMC} which includes the generic receptor model introduced in Section~\ref{subsec:genRec}.
The proposed receptor model is a generalization of the two-state \ac{AMPAR} model proposed in \cite{Lotter2021SaturatingModels} and we show in Section~\ref{sec:simulation} that the two-state \ac{AMPAR} model is indeed recovered from the three-state model. 

First, we reformulate the non-linear boundary condition \eqref{eq:bc:full} according to \cite[Eq.~(19)]{Lotter2021SaturatingModels} as follows
\begin{align}
   i'_{x'}(1,t') = \delta_{t'} o'(t') + \delta_{t'} d'(t') = \phi_\mathrm{i}'(t'),
   \label{eq:tf:0}
\end{align}
where boundary value $\phi_\mathrm{i}'$ is a placeholder yet to be defined in Section~\ref{subsec:incopRec}. 
The reformulated boundary condition \eqref{eq:tf:0}, together with \eqref{eq:dl:pde} and \eqref{eq:bc:left}, constitutes a dimensionless version of the SMC system investigated in \cite{Lotter2021SaturatingModels}. 
Hence, we employ a dimensionless version of the \ac{TFM} derived in \cite[Sec.~3]{Lotter2021SaturatingModels} as starting point\footnote{For a detailed description of how the TFM model is derived, we refer the reader to \cite[Sec.~3]{Lotter2021SaturatingModels}}. 
For the diffusion and degradation of NTs in the considered SMC system, we employ the following dimensionless discrete-time SSD consisting of $\mu = 0, \dots, Q-1$ parallel systems \cite[Eqs.~(30), (31), (39)]{Lotter2021SaturatingModels}
\begin{align}
    \bar{\bm{y}}'[k\!+\!1] \!&=\! \mathrm{e}^{\kap{e}'T'}\!\mathrm{e}^{\As'T'}\bar{\bm{y}}'[k] + T'\bar{\bm{f}}'[k\!+\!1] - T'\bar{\bm{\phi}}'[k\!+\!1], \label{eq:tf:1}\\
    c'[x',k] &= \bm{c}_1'^{\tra}(x')\bar{\bm{y}}'[k], \label{eq:tf:2}
\end{align}
with discrete-time step $k$, i.e., $t' = kT'$, normalized sampling interval $T' = T\nicefrac{D}{a^2}$, and transposition operator $(\cdot)\tran$. \textit{State equation} \eqref{eq:tf:1} describes the temporal evolution of the expansion coefficients $\bar{y}'_\mu$ in column vector $\bar{\bm{y}}' = \left(\bar{y}'_\mu \right)^{Q-1}_{\mu = 0}$ (see \cite[Eq.~(21), (28)]{Lotter2021SaturatingModels}). 
The eigenvalues of the \ac{NT} diffusion process, \mbox{$\lambda'_\mu = -\mu\pi$}, are arranged in matrix $\As' = \mathrm{diag}\{\lambda'_0, \dots, \lambda'_{Q-1}\}$. 
Vectors $\bar{\bm{f}}' = \left(f'_\mu\right)_{\mu = 0}^{Q-1}$ and $\bar{\bm{\phi}}' = \left(\bar{\phi}'_\mu\right)_{\mu = 0}^{Q-1}$ are the scalar-valued coefficients of the expansion of source function $s'$ and the placeholder boundary value $\phi_\mathrm{i}$ in \eqref{eq:tf:0}, respectively (see \cite[Eq.~(29)]{Lotter2021SaturatingModels}). \textit{Output equation} \eqref{eq:tf:2} recovers the concentration of NTs from the expansion coefficients $\bar{\bm{y}}$ by multiplication 
with row vector $\bm{c}_1'(x') = \left(\nicefrac{K'_\mu(x')}{N'_\mu}\right)_{\mu = 0}^{Q-1}$ comprising eigenfunctions $K'_\mu$ and scaling factor $N_\mu'$, cf. \cite[Eqs.~(26), (27), (32)]{Lotter2021SaturatingModels}.

\begin{figure}[t]
    \centering
    \includegraphics[width=0.8\linewidth]{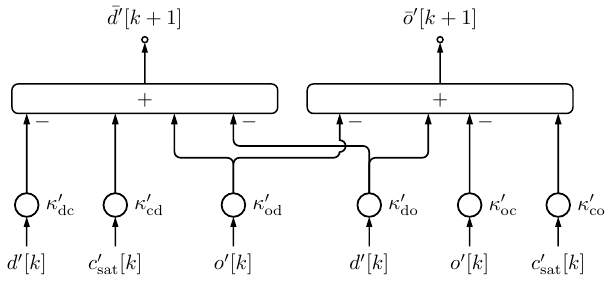}
    \vspace*{-0.3cm}
    \caption{Discrete-time generic receptor circuitry according to \eqref{eq:dt:r3}, \eqref{eq:dt:r4}.}
    \label{fig:receptorcircuit}
    \vspace*{-0.55cm}
\end{figure}

\vspace*{-0.2cm}
\subsection{Discrete-time Generic Receptor Model}
\vspace*{-0.1cm}
Before we incorporate the generic receptor model into SSD \eqref{eq:tf:1}, \eqref{eq:tf:2}, we transform boundary condition \eqref{eq:bc:full} into the discrete-time domain as follows 
\begin{align}
    i'_{x'}[1,k] = \bar{o}'[k] + \bar{d}'[k], 
    \label{eq:dt:r1}
\end{align}
where $\bar{o}'$ and $\bar{d}'$ accumulate to $o'$ and $d'$
\begin{align}
    &o'[k] = T\sum_{n=0}^{k} \bar{o}'[n], &d'[k] = T\sum_{n=0}^{k} \bar{d}'[n].
    \label{eq:dt:r2}
\end{align}
To resolve the implicit character of \eqref{eq:bc:open} and \eqref{eq:bc:desen} and to make the boundary condition computable, we introduce a delay of one time step $T'$ in the discrete-time domain, yielding
\begin{align}
    \bar{o}'[k+1] &= \kap{co}'c_\mathrm{sat}'[k] \!- \kap{oc}'o'[k] \!- \kap{od}'o'[k] \!+ \kap{do}'d'[k], \label{eq:dt:r3}\\
    \bar{d}'[k+1] &= \kap{cd}'c_\mathrm{sat}'[k] \!- \kap{dc}'d'[k] \!- \kap{do}'d'[k] \!+ \kap{od}'o'[k], \label{eq:dt:r4}
\end{align}
where $c'_\mathrm{sat}[k] = c_\mathrm{s}'[k]c'[1,k]$, with
\begin{align}
    c_s'[k] = 1 \!-\!\nicefrac{N_0}{C^*}\left(o'[k] + d'[k] \right)\!, \label{eq:dt:r5}
\end{align}
which models the receptor occupancy.
If $T'$, respectively $T$, is small compared to the binding rates, the additional delay is physically justified \cite{Lotter2021SaturatingModels}. Fig.~\ref{fig:receptorcircuit} shows the general receptor circuitry according to \eqref{eq:dt:r3} and \eqref{eq:dt:r4}. 


\subsection{Incorporation of the Generic Receptor Model}
\label{subsec:incopRec}

For incorporating the generic receptor model, we specialize the placeholder in \eqref{eq:tf:1} to discrete-time boundary condition \eqref{eq:dt:r1} as follows
\begin{align}
    \bar{\bm{\phi}}'[k+1] &= \tilde{\bm{c}}'_2(1)\phi_\mathrm{i}'[k+1] = \tilde{\bm{c}}_2'(1)i_x'[1,k+1]\nonumber\\ &=\tilde{\bm{c}}_2'(1)\left(\bar{o}'[k+1] + \bar{d}'[k+1]\right),
    \label{eq:ir:1}
\end{align}
where we exploited the relation between $\phi_\mathrm{i}'$ and its expansion coefficients $\bar{\bm{\phi}}'$ from \cite[Eq.~(29)]{Lotter2021SaturatingModels} and $\tilde{\bm{c}}_2'(1) = \left((-1)^{\mu}\right)_{\mu = 0}^{Q-1}$. 

\noindent Substituting \eqref{eq:dt:r3} and \eqref{eq:dt:r4} in \eqref{eq:ir:1} and exploiting \eqref{eq:tf:2} leads to 
\begin{align}
    \bar{\bm{\phi}}'[k+1] = \bm{\mathcal{K}}'&c_\mathrm{s}'[k]\left(\kap{co}' + \kap{cd}'\right) \bar{\bm{y}}'[k] \nonumber\\
    &- \tilde{\bm{c}}'_2(1)\left(\kap{oc}'o'[k] - \kap{dc}'d'[k]\right), 
    \label{eq:ir:2}
\end{align}
with $\bm{\mathcal{K}}' = \tilde{\bm{c}}'_2(1)\bm{c}_1'^{\tra}(1)$. 

Inserting \eqref{eq:ir:2} into state equation \eqref{eq:tf:1} leads to the modified state equation
\begin{align}
    \bar{\bm{y}}'&[k+1] \!=\! \left(\mathrm{e}^{-\kap{e}' T'}\mathrm{e}^{\As' T'} \!- T'\bm{\mathcal{K}}'c_\mathrm{s}'[k]\left(\kap{co}' + \kap{cd}'\right)\right)\bar{\bm{y}}'[k] \nonumber \\  
    &+ T'\tilde{\bm{c}}'_2(1)\left(\kap{oc}'o'[k] + \kap{dc}'d'[k]\right)
    	 + T'\bar{\bm{f}}'[k+1],
    \label{eq:ir:3}
\end{align}
which accounts for the generic receptor model at the post-synaptic cell and illustrates the influence of the different state transitions on the NT concentration (see Fig.~\ref{fig:receptorFull}). 
In particular, the binding processes decrease the NT concentration with rates $\kap{co}'$ and $\kap{cd}'$ in the presence of negative feedback due to receptor occupancy expressed by $c_\mathrm{s}'$ in \eqref{eq:dt:r5}.
The unbinding processes increase the NT concentration with rates $\kap{oc}'$ and $\kap{dc}'$. 
The transition between the open and desensitized states has no direct influence on the NT concentration in the channel, and therefore, in contrast to \eqref{eq:dt:r3} and \eqref{eq:dt:r4}, rates $\kap{od}'$ and $\kap{do}'$ do not occur in \eqref{eq:ir:3}. 


\subsection{Specialization to AMPAR and NMDAR}
\label{subsec:specmodel}

The kinetic schemes for \ac{AMPAR} and \ac{NMDAR} proposed in \cite{Destexhe1994SynthesisFormalism} and the dedicated receptor circuits can be obtained by a modification of the scheme in Fig.~\ref{fig:receptorFull} and the circuit in Fig.~\ref{fig:receptorcircuit}, respectively, which affects state equation \eqref{eq:ir:3}.
%
For example, to realize the three-state \ac{AMPAR}, $\kap{cd}$ and $\kap{do}$ have to be set to zero in \eqref{eq:dt:r3}, \eqref{eq:dt:r4} and \eqref{eq:ir:3}, yielding
\begin{align}
    \bar{\bm{y}}_{\mathrm{AM}}'&[k+1] \!=\! \left(\mathrm{e}^{-\kap{e}' T'}\!\mathrm{e}^{\As' T'} \!- T'\bm{\mathcal{K}}'c_\mathrm{s}'[k]\kap{co}' \right)\bar{\bm{y}}_{\mathrm{AM}}'[k] \nonumber \\  
    &+ T'\tilde{\bm{c}}'_2(1)\left(\kap{oc}'o'[k] + \kap{dc}'d'[k]\right)
    	 + T'\bar{\bm{f}}'[k+1],
    \label{eq:ir:4}
\end{align}
which contains the effect of saturation and desensitized states\footnote{We note that by setting $\kap{dc}'$ to zero, \cite[Eq.~(42)]{Lotter2021SaturatingModels} for a two-state AMPAR is recovered from \eqref{eq:ir:4} (in dimensionless form).}.  
A state equation similar to \eqref{eq:ir:4} for \ac{NMDAR} can be obtained by setting $\kap{co}'$ and $\kap{od}'$ to zero in \eqref{eq:dt:r3} and \eqref{eq:dt:r4}.

\section{Numerical Evaluation}
\label{sec:simulation}
In this section, numerical results obtained with the generic receptor model proposed in Sections~\ref{sec:system} and \ref{sec:tfm} are presented along with results from stochastic \acp{PBS}.
For the \acp{PBS}, we adopted the simulator design from \cite{Lotter2021SaturatingModels}.
The simulator features three-dimensional Brownian motion of \acp{NT} inside the synaptic cleft, reversible binding reactions of \acp{NT} with individual membrane-bound receptors, and homogeneous first-order degradation of \acp{NT}.
To incorporate the additional state transitions of the generic three-state receptor model, the respective additional first-order reactions were introduced into the simulator using Gillespie's algorithm \cite{Gillespie1977Exact}.
The results from \acp{PBS} were averaged over $10^3$ realizations.
The \ac{TFM} derived in Section~\ref{sec:tfm} was implemented in MATLAB and the corresponding source code is available online on \cite{code}.
\vspace*{-0.1cm}
\subsection{Default Parameter Values}
\vspace*{-0.1cm}
The three-state kinetic scheme in Section~\ref{sec:system} comprises first- and second-order chemical reactions.
In lack of experimental reference values for each of the possible state transitions in this model, the binding and unbinding rates of glutamate to and from \acp{AMPAR}, respectively, reported in \cite{holmes95,jonas93} were used as reference values for the second- and the first-order reaction rates, respectively. 
In the following, we denote these default parameter values for the rate constants in Fig.~\ref{fig:receptorFull} by a tilde, e.g., $\kco'=\kcot'$, with the dimensionless default values $\kcot'\!=\kcdt'\!=\mbox{$9.2\cdot\! 10^{-4}$}$ and $\koct'\!=\kdct'\!=\kodt'\!=\kdot'\!=5.2\cdot\! 10^{-3}$.
All other parameter values were taken from \cite[Table~1]{Lotter2021SaturatingModels}.

\begin{figure}[t]
    \centering
    \includegraphics[width=\linewidth]{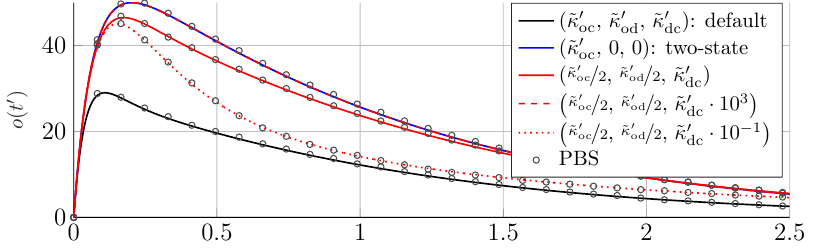}\\
    \includegraphics[width=\linewidth]{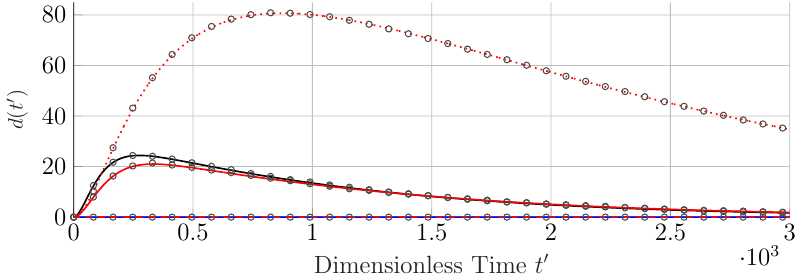}
    \vspace*{-0.6cm}
    \caption{Number of open ({\em top}) and desensitized ({\em bottom}) \acp{AMPAR} obtained from the SSD model in Section~\ref{sec:tfm} (solid lines) and PBS data (circle markers) for a single release of NTs and different transition rates $\left(\koc',\, \kod',\,\kdc'\right)$.}
    \label{fig:ampa_single_release}
    \vspace*{-0.6cm}
\end{figure}
\vspace*{-0.1cm}
\subsection{AMPAR Kinetic Scheme}
\begin{figure}[t]
    \centering
    \includegraphics[width=\linewidth]{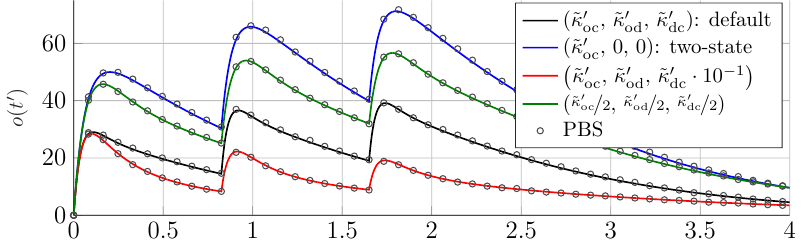}\\
    \includegraphics[width=\linewidth]{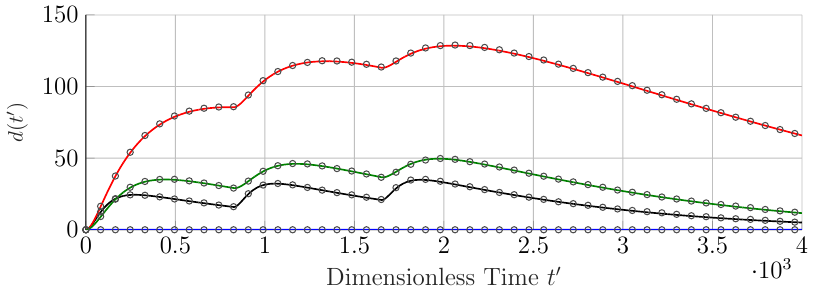}
    \vspace*{-0.6cm}
    \caption{Number of open ({\em top}) and desensitized ({\em bottom}) \acp{AMPAR} obtained from the SSD model in Section~\ref{sec:tfm} (solid lines) and PBS data (circle markers) for multiple releases of \acp{NT} and different transition rates $\left(\koc',\, \kod',\, \kdc'\right)$.}
    \label{fig:ampa_multiple_releases}
    \vspace*{-0.6cm}
\end{figure}

\vspace*{-0.1cm}
In this section, we consider the specialized kinetic scheme from Fig.~\ref{fig:receptorFull} proposed as model for \acp{AMPAR} in \cite{Destexhe1994SynthesisFormalism}. Figs.~\ref{fig:ampa_single_release} and \ref{fig:ampa_multiple_releases} show $o(t')$ and $d(t')$ resulting from single and multiple releases of \acp{NT} as obtained from the \ac{SSD} model proposed in Section~\ref{sec:tfm} and \ac{PBS} data, respectively, for different combinations of state transition rates.

\subsubsection{Single Release of NTs}

First, we note that the results from the proposed \ac{SSD} model and from \acp{PBS} match very well.
Next, as previously mentioned in Section~\ref{subsec:specmodel}, the proposed three-state \ac{AMPAR} model recovers the two-state \ac{AMPAR} model for $\kod'=0$ and $\kdc'=0$, cf.~Fig.~\ref{fig:ampa_single_release}.
Comparing the number of open receptors for the default parameter setting (solid black) with the two-state model (solid blue) in Fig.~\ref{fig:ampa_single_release}, we note that the presence of the desensitized state in the default setting decreases the number of open receptors as compared to the two-state model.
This is intuitive, since the possibility of transiting to the desensitized state in the three-state model decreases the average time a receptor remains in the open state as compared to the two-state model.
Next, we investigate what impact the desensitized state has on the number of open receptors besides decreasing the opening time of the receptors.
To this end, we halve $\kap{od}'$ and $\kap{oc}'$, such that the average receptor opening times in the three-state and the two-state models, respectively, are the same (solid red).
Fig.~\ref{fig:ampa_single_release} shows, that in this case, the two-state model is recovered if $\kdc'$ is large (dashed red).
Finally, we observe from Fig.~\ref{fig:ampa_single_release} that, as $\kdc'$ decreases, the peak number of open receptors is almost preserved while the signal decays faster in the three-state model compared to the two-state model (dotted red).
On the other hand, according to Fig.~\ref{fig:ampa_single_release}, the number of desensitized receptors increases as $\kdc'$ decreases.
In summary, as $\kdc'$ decreases, more \acp{NT} are buffered at desensitized receptors, decreasing the chance that freely diffusing bulk \acp{NT} encounter a free receptor.
This reduced binding probability of \acp{NT} to receptors accelerates the enzymatic degradation of \acp{NT} and leads to faster signal termination.
It is evident from this discussion that the desensitized state of the \acp{AMPAR} can play an important role in shaping synaptic signal transmission.

\subsubsection{Multiple Releases of NTs}

Now, multiple releases of \acp{NT} 
at $t'\!= 0,\,0.8\cdot \!10^{3}, \, 1.6\cdot \!10^{3}$
are considered.
From Fig.~\ref{fig:ampa_multiple_releases}, we observe that the number of open receptors in the three-state \ac{AMPAR} model (solid black) is less than in the two-state model (solid blue).
However, for the default parameter values, the qualitative behavior of the two models is similar in the sense that both models exhibit synaptic facilitation, i.e., the peak values of $o(t')$ increase from one release of \acp{NT} to the next.
If all transition rates of the three-state \ac{AMPAR} model are set to half of their respective default values to recover the mean receptor opening time of the two-state model (solid green), the synapse still shows short time facilitation with $o(t')$ being closer to the two-state model as compared to the default parameter setting \cite[Chap.~16]{MolToNet}.
However, when $\kdc'$ is decreased relative to the default parameter values ($\koc'$ and $\kod'$ unchanged), the peak numbers of open receptors show a decreasing trend from one \ac{NT} release to the next (solid red).
This decreasing trend is readily explained from the number of desensitized receptors $d(t')$ which is largest for the considered scenario as compared to the previously considered scenarios.
Namely, as $\kdc'$ decreases relative to $\koc'$ and $\kod'$, \acp{NT} are buffered at desensitized receptors, hence reducing the number of unbound receptors exposed to subsequently released \acp{NT}.
In summary, the observations made in Fig.~\ref{fig:ampa_multiple_releases} indicate that the desensitized state can even impact the qualitative behavior of synapses over multiple \ac{NT} releases.
This suggests that the desensitized state can be critical in determining whether a synapse shows synaptic facilitation or depression which in turn is critical for assessing the functional role of the synapse \cite[Chap.~16]{MolToNet}.

\vspace*{-0.1cm}
\subsection{NMDAR Kinetic Scheme}
\begin{figure}[t]
    \centering
    \includegraphics[width=\linewidth]{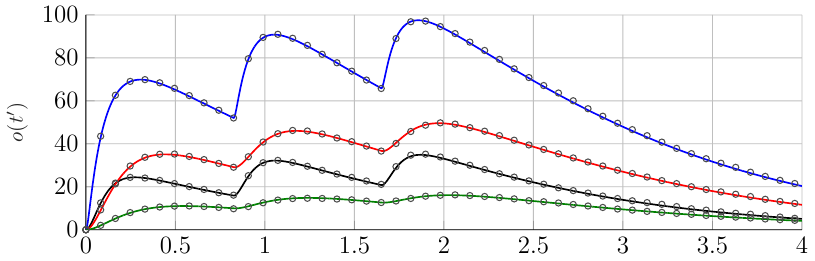}\\
    \includegraphics[width=\linewidth]{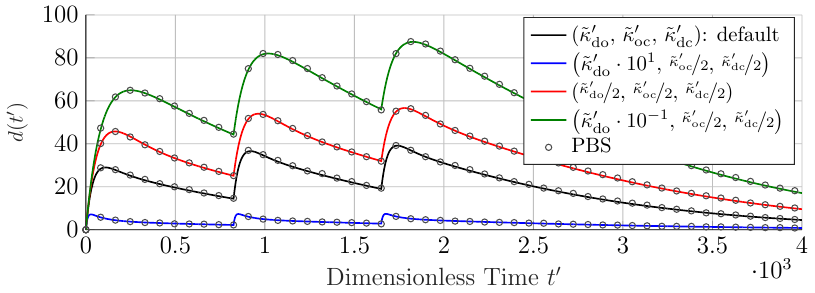}
    \vspace*{-0.7cm}
    \caption{Number of open ({\em top}) and desensitized ({\em bottom}) \acp{NMDAR} obtained with the \ac{SSD} model in Section~\ref{sec:tfm} (solid lines) and \ac{PBS} data (circle markers) for multiple releases of \acp{NT} and different transition rates $\left(\kdo',\, \koc',\, \kdc'\right)$. }
    \label{fig:nmda_multiple_releases}
    \vspace*{-0.7cm}
\end{figure}

\vspace*{-0.1cm}
In this section, the three-state kinetic model for \acp{NMDAR} presented in Fig.~\ref{fig:receptorFull} is investigated.
Fig.~\ref{fig:nmda_multiple_releases} shows the activation of \acp{NMDAR} after multiple releases of \acp{NT} as predicted by the \ac{SSD} model proposed in Section~\ref{sec:tfm} and data obtained with \acp{PBS}.
First, comparing $d(t')$ and $o(t')$ in Fig.~\ref{fig:nmda_multiple_releases}, we observe that for each of the considered three releases of \acp{NT}, after an \ac{NT} release, the number of desensitized receptors assumes a local maximum {\em before} the number of open receptors assumes a local maximum.
For the default parameter setting (solid black), for example, $d(t')$ peaks for the first time at $t' \approx 0.1 \times 10^{3}$, while the first peak of $o(t')$ is observed at $t' \approx 0.25 \times 10^{3}$.
This is a key difference to the \ac{AMPAR} model studied in the previous section and results from the kinetic scheme of \ac{NMDAR} receptors, which forces the receptors to transition directly into the desensitized state upon \ac{NT} binding.
Hence, two state transitions are needed for \acp{NMDAR} to reach the open state from the closed state, while one single state transition is sufficient for \acp{AMPAR} (see Fig.~\ref{fig:receptorFull}).
The delayed response of \acp{NMDAR} compared to \acp{AMPAR} is in agreement with experimental results that suggest that \acp{NMDAR} mediate the slow-rising component of the electrical downstream signal at the postsynaptic cell while \acp{AMPAR} contribute mainly the fast-rising component 
\cite[Chap.~16]{MolToNet}.
Next, we consider the case when all first-order reaction rates in the \ac{NMDAR} model are reduced to half of their default values such that all first-order state transitions occur less frequently, but with the same relative probabilities as in the default setting (solid red).
From Fig.~\ref{fig:nmda_multiple_releases}, we observe that in this case the numbers of desensitized and open receptors increase both.
This observation is expected, since more \acp{NT} accumulate at the receptors due to the reduced reaction rates.
Finally, we study the impact of the transition from desensitized to open state, $\kdo'$.
From Fig.~\ref{fig:nmda_multiple_releases}, we observe that increasing $\kdo'$ leads to an acceleration in the receptor opening of the \acp{NMDAR} compared to the previously considered case, while the number of desensitized \acp{NMDAR} decreases (solid blue).
On the other hand, when $\kdo'$ is decreased, the opening of \acp{NMDAR} occurs more slowly while the number of desensitized \acp{NMDAR} increases as compared to the previously considered cases (solid green).
These observations agree with the fact that \ac{NMDAR} responses can be relevant for long-term signaling which accumulates information over multiple synaptic transmission events 
\cite[Chap.~16]{MolToNet}.

\vspace*{-0.1cm}
\section{Conclusion}
\label{sec:conclusion}
In this paper, a novel deterministic signal model for \ac{SMC} was proposed.
The model encompasses the main types of postsynaptic receptors and allows to study how these transduce the chemical synaptic signal.
By applying dimensional analysis and conducting \acp{PBS}, the generality of the proposed model as well as its accuracy have been verified.
Numerical results show that the proposed model allows for analyses that exceed the capabilities of the commonly used two-state receptor models.

In future work, we plan to extend the proposed model to account for the competition of different receptor types for \acp{NT} in the same synapse.
Apart from \ac{SMC}, it would be interesting to apply the proposed model for the design of synthetic \ac{MC} systems based on reactive \acp{Rx}.

\IEEEpeerreviewmaketitle

\bibliographystyle{IEEEtran}
{\footnotesize
    \vspace*{-0.1cm}
	\bibliography{./IEEEabrv,./references_fin}}

\end{document}